\documentclass[twocolumn,showpacs,aps,groupedaddress,prd,eqsecnum,epsf]{revtex4}

\usepackage{amsfonts,latexsym,eucal}
\usepackage[dvips]{graphicx}

\newcommand{\dalm}{\kern1pt\vbox{\hrule height 0.9pt\hbox{\vrule width 0.9pt
\hskip 2.5pt\vbox{\vskip 5.5pt}\hskip 3pt\vrule width 0.3pt}\hrule height 0.3pt}\kern1pt}

\usepackage{epsfig}

\begin{document}
% \draft

\thispagestyle{empty}

%%    -------
\title{Generic features of Einstein-Aether black holes}
\author{Takashi Tamaki}
\email{tamaki@gravity.phys.waseda.ac.jp}
%%    -------
\author{Umpei Miyamoto}
\email{umpei@gravity.phys.waseda.ac.jp}
\address{Department of Physics, Waseda University, Okubo 3-4-1, Tokyo 169-8555, Japan}
%%    -------

%\date{\today}

\begin{abstract}
We reconsider spherically symmetric black hole solutions in Einstein-Aether 
theory with the condition that this theory has identical PPN parameters as those for general relativity, 
which is the main difference from the previous research. 
In contrast with previous study, we allow superluminal propagation of 
a spin-$0$ Aether-gravity wave mode. 
As a result, we obtain black holes having a spin-$0$ ``horizon" inside an event horizon. 
We allow a singularity at a spin-$0$ ``horizon" since it is concealed by the event horizon. 
If we allow such a configuration, 
the kinetic term of the Aether field can be large enough for 
black holes to be significantly different from Schwarzschild black holes 
with respect to ADM mass, 
innermost stable circular orbit, Hawking temperature, and so on. 
We also discuss whether or not the above features can be seen in more generic 
vector-tensor theories. 
\end{abstract}

\maketitle

%%%%%%%%%%%%%%%%%%%%%%
\section{Introduction}
%%%%%%%%%%%%%%%%%%%%%%

Identifying the contents of dark energy and dark matter (DE/DM) is one of the most important 
subjects in cosmology. It is frequently argued that gravitational theories are an alternative 
to DE/DM. Recently, tensor-vector-scalar (TeVeS) theories have attracted much 
attention since they do not only explain galaxy rotation curves but also satisfy many constraints 
from solar experiments~\cite{Bekenstein}. Although a deficency in explaining the mismatch between 
luminous and dynamical masses in clusters of galaxies by TeVeS has been pointed out~\cite{Clowe}, 
resolution of this problem by considering a generalized vector-tensor theory has also been 
reported~\cite{Zlosnik}. Moreover, these vector fields might explain an accelerated expansion of 
the universe~\cite{Zlosnik,Tartaglia} and might be important in inflationary 
scenarios~\cite{Lim,Jiro}. 
The origin of such a vector field is argued in \cite{Koste}.

However, it is nontrivial whether or not these theories satisfy the constraints by 
strong gravity tests. 
Notice the result for scalar-tensor theories 
where compact objects have strong deviations from those in general relativity (GR) even in the 
cases that satisfy weak field tests~\cite{Damour}. 
%These researches including vector fields are difficult to perform in general. 
%In TeVeS, researches are still limited to special cases such as analyzed in \cite{Giannios}. 
To study vector fields in a general form is difficult. For example, 
results in TeVeS are still limited to special cases such as \cite{Giannios}. 
Thus, as a first step, it is important to investigate a simplified model which is tractable and 
instructive for general cases. 
One such useful model would be Einstein-Aether (EA) theory~\cite{start},
where all parameterized post-Newtonian (PPN) parameters~\cite{Will}
can be the same as those in GR~\cite{PPN}. 
EA theory is a vector-tensor theory, and TeVeS can be written as 
a vector-tensor theory which is the extension of EA theory~\cite{Zlosnik2}.   
In EA theory, strong gravitational cases including black 
holes have been analyzed to some extent~\cite{SS,BHs,QNM,Garfinkle,NS}. 

%Nevertheless, since analysis of black holes is limited the case in which 
%the event horizon coincides with the spin-$0$ horizon~\cite{BHs}, this case does not 
%necessarily satisfy weak fields tests. Thus, it is interesting to ask whether or not 
%significant differences from Schwarzschild black hole appear when weak fields tests 
%are satisfied. For this reason, we argue black holes 
%with the case that EA theory has identical PPN parameters in GR. 

Nevertheless, the analysis of black holes has been limited to the case in which 
the event horizon coincides with the spin-$0$ horizon~\cite{BHs}, and this case does not 
necessarily satisfy weak fields tests. Thus, it is interesting to ask whether or not 
significant differences from the Schwarzschild black hole appear when weak fields tests 
are satisfied. For this reason, we argue black holes 
with the case in which the EA theory has identical PPN parameters as in GR.

This paper is organized as follows. In Sec.~II, we explain EA theory and summarize 
constraints located by previous research. In Sec.~III, we mention our method of 
analyzing black holes. In Sec.~IV, we show the results and compare them with the Schwarzschild 
black hole. In Sec.~V, consequences and future subjects are discussed. 
In Appendix A, we summarize basic equations. We use units in which $c=1$ 
and follow the sign conventions of Misner, Thorne, and
Wheeler~\cite{Misner:1974qy}, e.g., $(-,+,+,+)$ for metrics. 

%%%%%%%%%%%%%%%%%%%%%%%%%%%%%%%%%%%%%%%
%%%%%%%%%%%%%%%%%%%%%%%%%%%%%%%%%%%%%%%
\section{Einstein-Aether theory}
%%%%%%%%%%%%%%%%%%%%%%%%%%%%%%%%%%%%%%%
%%%%%%%%%%%%%%%%%%%%%%%%%%%%%%%%%%%%%%%

%%%%%%%%%%%%
\subsection{The action and basic equations}
%%%%%%%%%%%%
We consider the following action~\cite{Garfinkle}:
\begin{eqnarray}
\hspace{-10mm}&&
	I
	=
	\frac{ 1 }{ 16 \pi G } \int
	d^4 x \sqrt{-g} \; \mathcal{L}\ ,
	\\
\hspace{-10mm}&&
	\mathcal{L}=
	  R
	- K^{ ab }_{ \;\;\;cd } \nabla_a u^c \nabla_b u^d
	+ \lambda ( u^2+1 )\ , 
	\\
\hspace{-10mm}&&
	K^{ ab }_{ \;\;\; cd }
	:=
	  c_1 g^{ ab } g_{ cd }
	+ c_2 \delta^a_c   \delta^b_d
	+ c_3 \delta^a_d   \delta^b_c
	- c_4 u^a    u^b   g_{ cd }\ ,
\end{eqnarray}
where
%$\mathcal{R}$ is the Ricci scalar curvature.
$u^{a}$ is a vector field 
and $ u^2 := u^a u_a$. $c_i$ ($i=1,2,3,4$) are theoretical 
parameters in EA theory. $\lambda$ is a Lagrange multiplier ensuring the vector 
field $u^a$ to be unit timelike vector everywhere. 

Varying this action with respect to $\lambda$ and $u^a$, we have
\begin{eqnarray}
	&&
	u^2+1 = 0\ ,
	\label{eq:var-lambda}
	\\
	&&
	c_4 \dot{u}^m \nabla_a u_m+\nabla_m J^m_{\;\;a}
	+ \lambda u_a
	=0\ ,
	\label{eq:var-u}
\end{eqnarray}
where
\begin{eqnarray}
	&&
	J^{a}_{\;\;m}
	:=
	K^{ab}_{\;\;\;\;mn} \nabla_{b} u^{n}\ ,
	\\
	&&
	\dot{u}^b
	:=
	u^a \nabla_a u^b\ .
\end{eqnarray}
Multiplying Eq.~(\ref{eq:var-u}) by $u_a$, we have
\begin{eqnarray}
	\lambda 
	= c_4 \dot{u}^2+u^a \nabla_{m} J^{m}_{\;\;a}\ .
	\label{eq:lambda}
\end{eqnarray}
Varying the action with respect to the metric, we have
\begin{eqnarray}
	G_{ ab }&=&\nabla_m
	\left[
		J^{m}_{\;\;(a} u_{b)}-J_{(a}^{\;\;m} u_{b)}+J_{(ab)} u^m
	\right]  \nonumber  \\
	&&+ c_1
	\left(
		\nabla_a u_m \nabla_b u^m-\nabla_m u_a \nabla^m u_b
	\right)  \nonumber  \\
	&&+ c_4 \dot{u}_a \dot{u}_b+\lambda u_a u_b
	- \frac{1}{2} g_{ab} \mathcal{L}_{\mathrm{u}}\ ,
	\label{eq:Einstein}
\end{eqnarray}
where 
%%%%%%%%%%%%%%%%%%
\begin{eqnarray}
	\mathcal{L}_{\mathrm{u}}
	:=
	K^{ab}_{\;\;\;\;cd} \nabla_a u^c \nabla_b u^d\ .
\end{eqnarray}
%%%%%%%%%%%%%%%%%%

%%%%%%%%%%%%%%%%%%
%%%%%%%%%%%%%%%%%%
\subsection{Present Constraints in EA Theory}
%%%%%%%%%%%%%%%%%%
%%%%%%%%%%%%%%%%%%

If we assume the weak field and slow-motion limits in EA theory~\cite{PPN}, 
we have to take Newton's gravitational constant as 
%%%%%%%%%%%%%%%%%%
\begin{eqnarray}
G_{\rm N}=\left(1-\frac{c_{1}+c_{4}}{2}\right)^{-1}G\ ,\label{Newton}
\end{eqnarray}
%%%%%%%%%%%%%%%%%%
to reproduce Newtonian gravity correctly.
For all the PPN parameters to coincide with those in GR,
we have 
%%%%%%%%%%%%%%%%%%
\begin{eqnarray}
c_{2}=\frac{-2c_{1}^{2}-c_{1}c_{3}+c_{3}^{2}}{3c_{1}}\ ,\ \ \ 
c_{4}=-\frac{c_{3}^{2}}{c_{1}}\ . \label{c2-2}
\end{eqnarray}
%%%%%%%%%%%%%%%%%%

If we assume Friedmann-Robertson-Walker (FRW) space-time and the Aether is aligned with a cosmological 
rest frame, the cosmological gravitational constant is given by~\cite{FRW} 
%%%%%%%%%%%%%%%%%%
\begin{eqnarray}
G_{\rm cosmo}=G\left( 1+\frac{c_{+}+3c_{2}}{2}\right)^{-1}\ , 
\end{eqnarray}
%%%%%%%%%%%%%%%%%%
where $ c_{+} := c_{1}+c_{3} $. Using primordial $^{4}\textrm{He}$ abundance, we have 
%%%%%%%%%%%%%%%%
\begin{eqnarray}
|G_{\rm cosmo}/G_{N}-1|<1/8\ .  \label{abundance}
\end{eqnarray}
%%%%%%%%%%%%%%%

%If we consider neutron stars~\cite{NS}, we can restrict $ c_{1}+c_{4} \leq 0.5\sim 1.6 $ 
%(which depends on equation of states) from 
%the maximum mass neutron star $\sim 2M_{\odot}$~\cite{Quain,Nice}. 
From the maximum mass of neutron stars $\sim 2M_{\odot}$~\cite{Quain,Nice},
we have $ c_{1}+c_{4} \leq 0.5\sim 1.6 $, depending on EOS~\cite{NS}.

In \cite{wave}, the sound modes are analyzed by expanding the metric and the Aether 
around the Minkowski metric. As in the case in GR, we have two spin-$2$ modes. 
As peculiar to EA theory, there are three wave modes. Two correspond to a 
transverse spin-$1$ mode, and one corresponds to a longitudinal spin-$0$ mode. 
The squared speeds of them are summarized as 
%%%%%%%%%%%%%%%%
\begin{eqnarray}
(s_{0})^{2}&=&\frac{c_{13}}{3(c_{1}-c_{3})(1-c_{13})}\ , \\
(s_{1})^{2}&=&\frac{c_{1}(2c_{1}-c_{1}^{2}+c_{3}^{2})}{2(1-c_{13})c_{13}(c_{1}-c_{3})}\ , \\
(s_{2})^{2}&=&\frac{1}{1-c_{13}}\ ,
\end{eqnarray}
%%%%%%%%%%%%%%%
where we eliminate $c_{2}$ and $c_{4}$ with Eq.~(\ref{c2-2}). 

For these sound velocities to be equal to or larger than 
the photon velocity, or, to ensure stability against linear perturbation in 
Minkowski (or FRW) background and linearized energy positivity, 
we have~\cite{wave,Lim,Elliott,Eling} 
%%%%%%%%%%%%%%%%
\begin{eqnarray}
0&<&c_{+}<1\ ,\nonumber  \\
0&<&c_{-}:=c_{1}-c_{3}<\frac{c_{+}}{ 3(1-c_{+}) }\ . \label{sonic}   
\end{eqnarray}
%%%%%%%%%%%%%%%
Radiation damping was also analyzed in \cite{damping,Foster}, which almost restricts 
$c_{+}$ as a function of $c_{-}$ based on the observation of, say, B1913+16~\cite{Stairs}. 
%Thus, the EA theory is highly restricted by these constraints. 

%%%%%%%%%%%%%%%%%%%%%%%%%%%%%%%%%%%%%%%%%%%%%%%%%%%%%%%%%%%%%%%%%%%%%%%%
\section{Analysis in a single-null coordinate system}
%%%%%%%%%%%%%%%%%%%%%%%%%%%%%%%%%%%%%%%%%%%%%%%%%%%%%%%%%%%%%%%%%%%%%%%%

Our purpose in investigating black holes in EA theory is 
not to give a further restriction but to understand generic features of 
vector-tensor theories under the condition that weak gravity tests are satisfied. 
This is the main difference from the previous research~\cite{BHs}, which investigates 
black holes with the parameters~\cite{foot-replace} 
%%%%%%%%%%%%%%%%
\begin{eqnarray}
c_{2}=\frac{-(c_{1}+c_{3})^{2}(c_{1}-c_{4})-2(c_{3}+c_{4})}{
(c_{1}-c_{4})(3c_{3}+4c_{4}-c_{1})+2 }
\end{eqnarray}
%%%%%%%%%%%%%%%%
and $c_{3}=-c_{4}$, or $c_{3}=-2c_{4}+c_{1}$, or $c_{3}=-c_{1}$. In these parameters, 
qualitative differences from Schwarzschild black holes have been shown. 
It is nontrivial whether or not this is true even for the case which satisfies 
weak gravity tests. 

From this point of view, we take the following strategy. 
(i) We assume (\ref{c2-2}) since the constraints by the solar experiments are severe. 
(ii) We assume (\ref{sonic}). Otherwise, a naked singularity appears outside the 
event horizon in general. 
As for other constraints, notice that (\ref{abundance}) is satisfied if (\ref{c2-2}) is satisfied. 
Constraints from neutron stars and from radiation damping are related to 
strong gravity tests at least partially. For the above reasons, we do not 
impose these constraints. 
Thus, we have two theoretical parameters $(c_{+},c_{-})$ with the condition (\ref{sonic}). 

We write a static and spherically symmetric line 
element in a single null coordinate system as,
%%%%%%%%%%%%%%%%
\begin{eqnarray}
	ds^2
	=
	- N(r) dv^2
	+ 2 B(r) dv dr
	+ r^2 d\Omega^{2}\ . \label{eq:single-null}
\end{eqnarray}
%%%%%%%%%%%%%%%%
In this coordinate, the vector field takes the form of
\begin{eqnarray}
	u
	=
	  a(r) \partial_v
	+ b(r) \partial_r\ . \label{vector-setup}
\end{eqnarray}
$b(r)\neq 0$ means that the Aether is not aligned with the timelike Killing field, 
which is inevitable because of the event horizon. 
From Eq.~(\ref{eq:var-lambda}), 
\begin{eqnarray}
	-N a^2+2 B a b = -1\ . \label{unit}
\end{eqnarray}

We can eliminate $\lambda$ with Eq.~(\ref{eq:lambda}). 
Then, from the Einstein and Aether equations, we obtain 
basic equations, which can be written schematically as
%%%%%%%%%%%%%%%%%%%%
\begin{eqnarray}
	N'&=&f_{1}(B,N,a,a')\ ,  \label{basiceq1}  \\
	B'&=&f_{2}(B,N,a,a')\ ,  \label{basiceq2}  \\
   a''&=&f_{3}(B,N,a,a')\ ,  \label{basiceq3}
\end{eqnarray}
%%%%%%%%%%%%%%%%%%%%
where the prime denotes the derivative with respect to $r$.
Here, we have eliminated $b$ with Eq.~(\ref{unit}). 
The explicit form is summarized in Appendix A.  

The boundary condition at the horizon $r_{\rm h}$ is $N(r_{\rm h})=0$.
We set $B(r_{\rm h})=1$. We can also set $ r_{\rm h}=1 $ since there is no 
scale in the present theory.
In this sense, it is assumed that the area coordinate $r$ 
is normalized by the horizon radius below. 

If we use a rescaling freedom of $v$ as $dv'=B(\infty ) dv$, 
the asymptotic form of the metric is written as 
%%%%%%%%%%%%%%%%
\begin{eqnarray}
	ds^2
	=
	- \frac{N(\infty )}{B(\infty )^{2}} dv'^2
	+ 2 dv' dr
	+ r^2 d\Omega^{2}\ . \label{eq:single-null3}
\end{eqnarray}
%%%%%%%%%%%%%%%%
Thus, the boundary condition at spatial infinity for the asymptotic flatness is 
%%%%%%%%%%%%%%%%%%%%
\begin{eqnarray}
	N(\infty )
	=
	B(\infty )^{2}\ .
	\label{flat}
\end{eqnarray}
%%%%%%%%%%%%%%%%%%%%
We should require
%%%%%%%%%%%%%%%%%%%%
\begin{eqnarray}
	b(\infty )=0\ ,
	\label{flat2}
\end{eqnarray}
%%%%%%%%%%%%%%%%%%%%
for the Aether to be aligned with the timelike Killing field.  
Then, by Eq.~(\ref{unit}), we have 
%%%%%%%%%%%%%%%%%%%%
\begin{eqnarray}
	a(\infty )=\frac{1}{B(\infty )}\ .
	\label{flat3}
\end{eqnarray}
%%%%%%%%%%%%%%%%%%%%
We can determine the pair of $a_{\rm h}:=a(r_{\rm h})$ and $a'_{\rm h}:=a'(r_{\rm h})$ as 
shooting parameters, one of which is fixed by (\ref{flat3}). 
Thus, there remains one freedom. Fixing this freedom is done as follows. 

Even in the spherically symmetric case, there is a spin-$0$ mode. 
Then, we can define the effective metric for a spin-$0$ mode as 
%%%%%%%%%%%%%%%%%%%%
\begin{eqnarray}
g_{ab}^{(0)}=g_{ab}-[(s_{0})^{2}-1]u_{a}u_{b}	\ .
\end{eqnarray}
%%%%%%%%%%%%%%%%%%%%
We call the horizon associated with this metric as the spin-$0$ horizon. 
The freedom mentioned above is fixed by the requirement that the regularity at 
the spin-$0$ horizon which is inside the event horizon. 

However, since the asymptotic observer is insensitive to the regularity at the spin-$0$ 
horizon, we permit the singularity at the spin-$0$ horizon. 
For this reason, we leave one freedom. 
In concrete terms, we obtain $a_{\rm h}$ iteratively for some $a'_{\rm h}$, which is regarded 
as a free parameter. 
We use the Bulirsch-Stoer method in our numerics~\cite{Press}.

%%%%%%%%%%%%%%%%%%%%
\begin{figure}[htbp]
\psfig{file=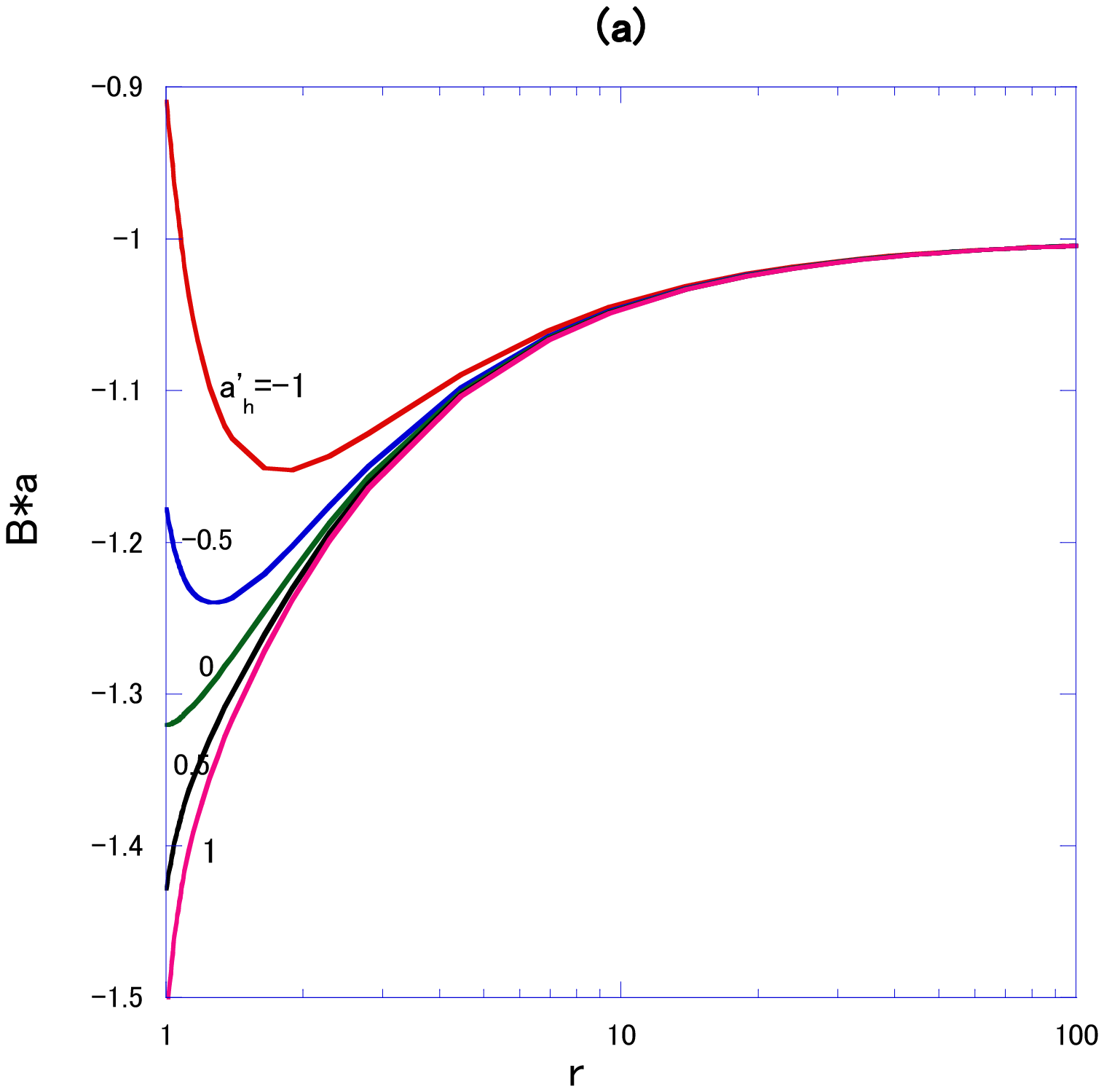,width=3in}
\psfig{file=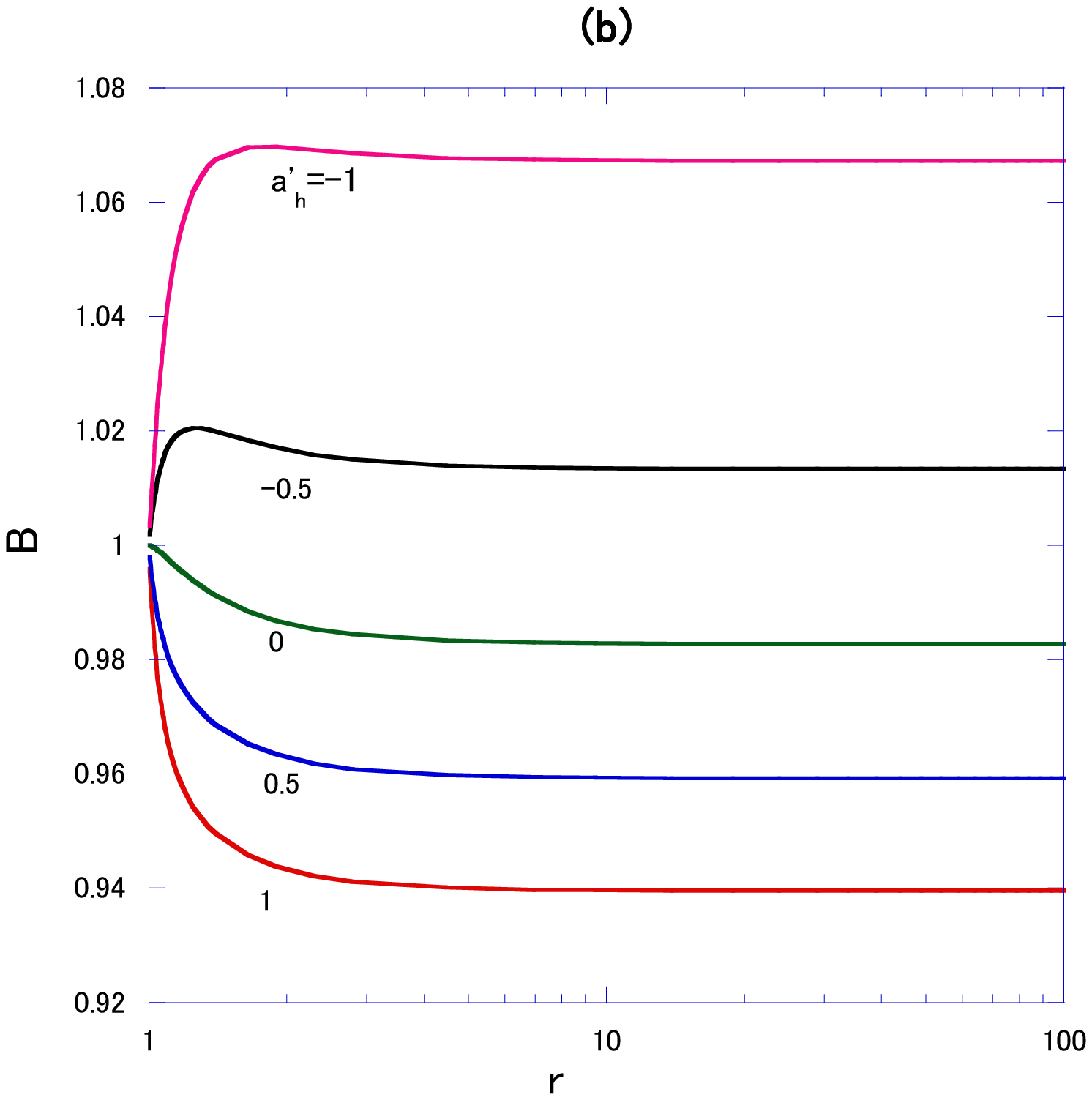,width=3in}
\psfig{file=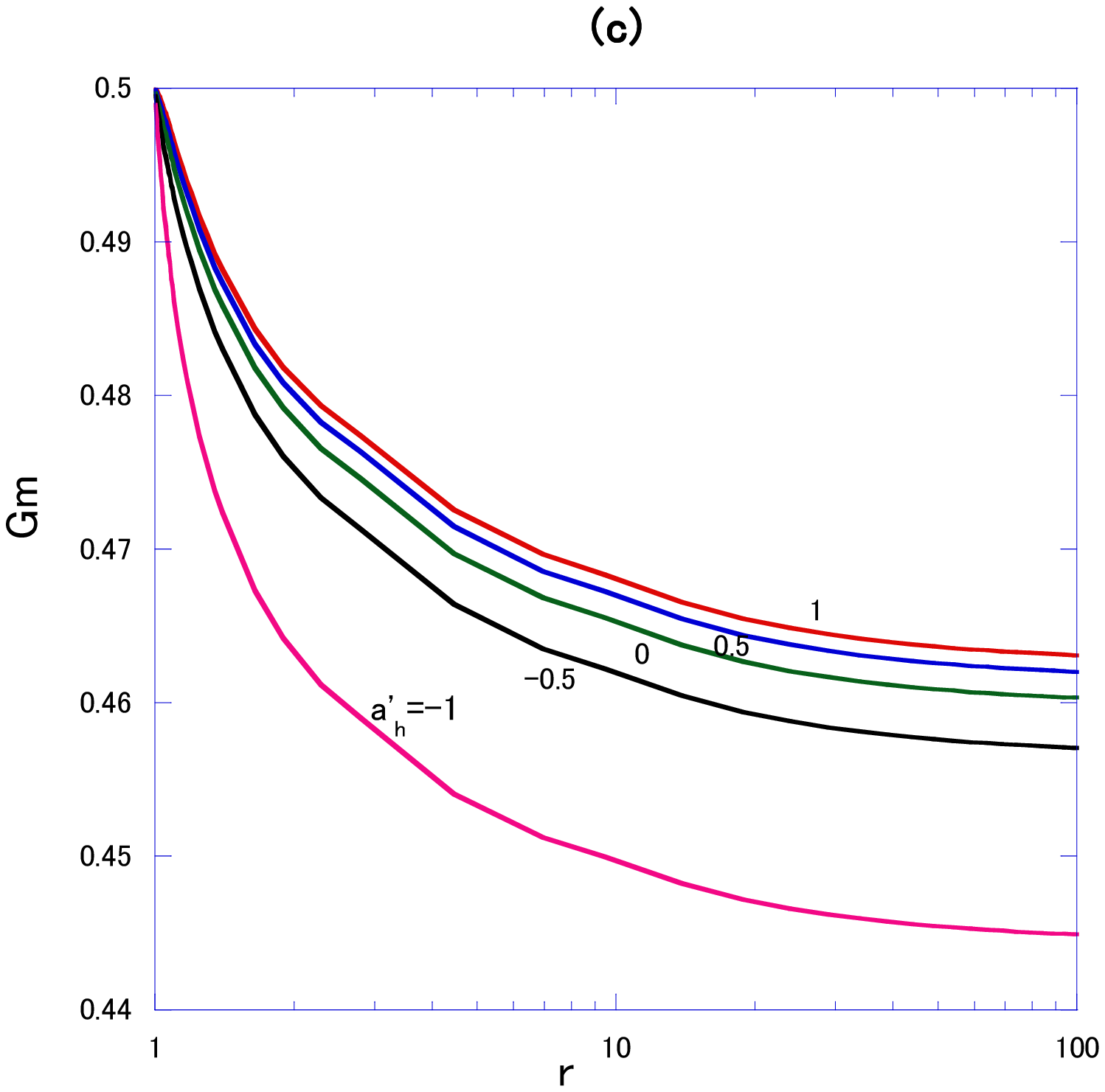,width=3in}
\caption{Field configurations for $c_{+}=0.4$ and $c_{-}=0.1$.
Denoted numbers in each figure, ranging from $-1$ to $1$,
represent the values of $a'_{\rm h}$. We normalize the quantities $Gm$ and $r$ 
by the horizon radius $r_{h}$. The solution with the smallest $a'_{\rm h}$ has 
largest deviation from a Schwarzschild black hole. 
\label{field} }
\end{figure}
%%%%%%%%%%%%%%%%%%%%%
%%%%%%%%%%%%%%%%%%%%
\begin{figure}[htbp]
\psfig{file=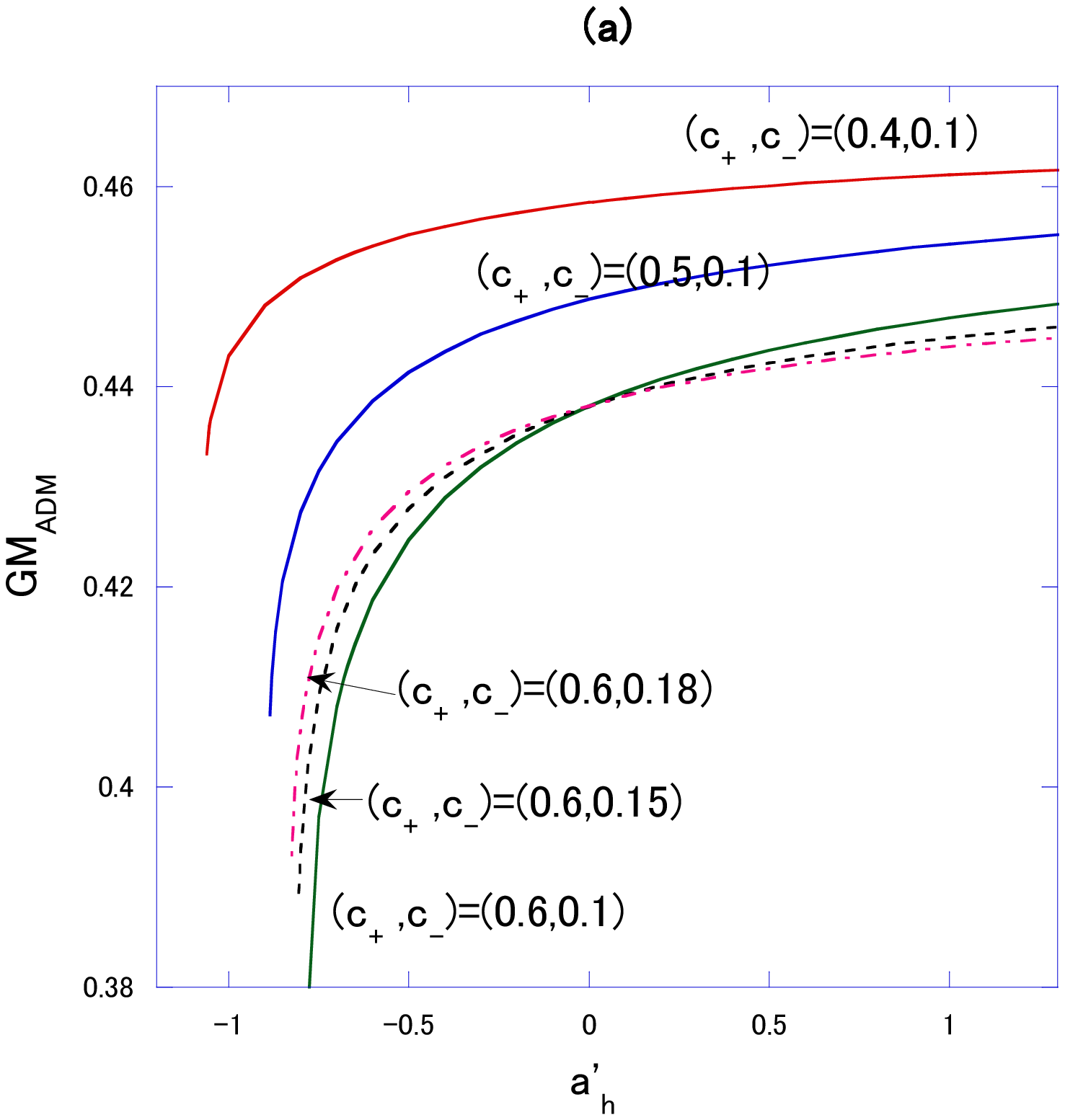,width=3in}
\psfig{file=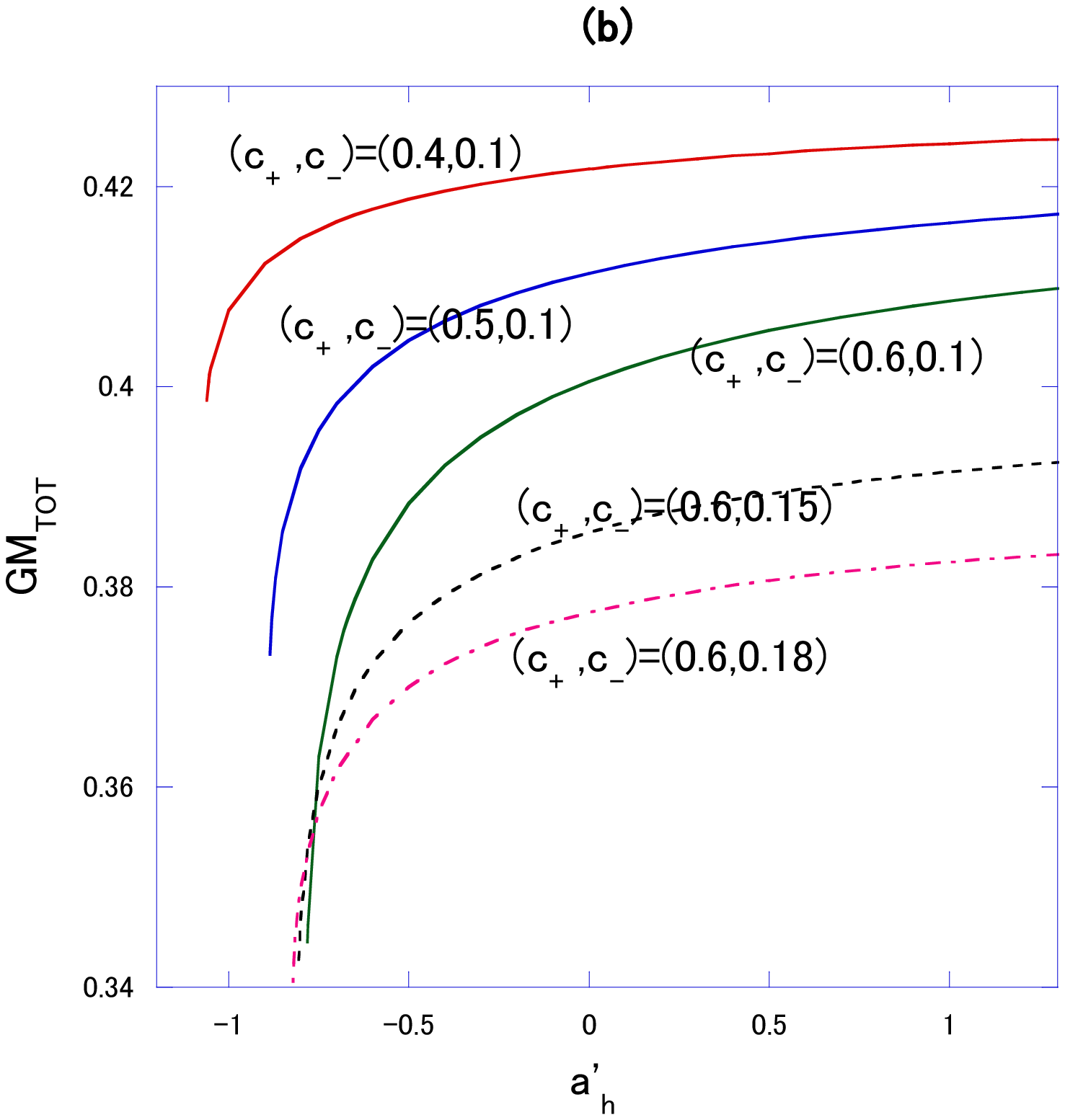,width=3in}
\caption{ (a) $a'_{\rm h}$ v.s. $ GM_{\rm ADM} $ and (b) $a'_{\rm h}$ v.s. $ GM_{\rm TOT} $ for 
several sets of $c_{+}$ and $c_{-}$. Physical quantities are normalized 
by the horizon radius $r_{\rm h}$. Notice that there is a lower limit $a'_{\rm h,crit}$
below which there is no regular solution. Near $a'_{\rm h,crit}$, $ GM_{\rm ADM} $ and 
$GM_{\rm TOT} $ depend on $a'_{\rm h}$ remarkably. 
\label{a'-M} }
\end{figure}
%%%%%%%%%%%%%%%%%%%
%%%%%%%%%%%%%%%%%%%%%%%%%%%%%%%%%
%%%%%%%%%%%%%%%%%%%%%%%%%%%%%%%%%
\section{properties of solutions}
%%%%%%%%%%%%%%%%%%%%%%%%%%%%%%%%%
%%%%%%%%%%%%%%%%%%%%%%%%%%%%%%%%%
\subsection{Mass and Hawking temperature of EA black hole}

We show several asymptotically flat solutions in Figs.~\ref{field} (a)-(c) 
for $ c_{+}=  0.4 $ and $ c_{-} = 0.1 $. In the figures, we have selected five solutions. 
The differences of these solutions are the changing boundary value $a'_{\rm h}$, ranging from $-1$ to $1$, 
as denoted in the figures. 
Figure~\ref{field} (a) shows that we can determine an $ a_{\rm h} $ 
that satisfies the asymptotic condition (\ref{flat3}) for various values of $a'_{\rm h}$. 
We also show $B(r)$ in Fig.~\ref{field} (b). Since $B(r) = \textrm{const}. = 1$ for a 
Schwarzschild black hole, 
it indicates that there are differences in physical quantities from those for Schwarzschild black holes.
Figure~\ref{field} (c) shows a ``mass" function. 
In AE theory, it is important to distinguish different notions of mass. 
If we define the mass function $m(r)$ by 
%%%%%%%%%%%%%%%%%%%%
\begin{eqnarray}
	m(r) := \frac{ r }{ 2G } \left(1-\frac{N}{B^{2}}\right)\ , 
\end{eqnarray}
%%%%%%%%%%%%%%%%%%%%
we can interpret $m(\infty )$ as ADM mass $M_{\rm ADM}$. 
As we can see, $ m(r) $ monotonically decreases. 
Our calculation suggests that this is generic. This is not surprising since 
energy conditions are not necessarily satisfied in EA theory~\cite{Eling}. 

Since Figs.~\ref{field} show that the deviation from the Schwarzschild black hole
is largest for the smallest value of $a'_{\rm h}$,
it is natural to ask whether or not there is a lower limit $a'_{\rm h,crit}$
below which there is no regular solution. 
We show the relation $a'_{\rm h}$ and $ M_{\rm ADM} $ for various values of $c_{+}$ and 
$c_{-}$ in Fig.~\ref{a'-M} (a). Typically, $ M_{\rm ADM} $ is smaller than that of a 
Schwarzschild black hole by about $10\%$, which is consistent with the result in \cite{BHs}. 
Here, we obtain $a_{\rm h}$ iteratively to satisfy Eq.~(\ref{flat}) for each $a'_{\rm h}$.
For $ a'_{\rm h} < a'_{\rm h,crit} $,
we cannot find an appropriate value of $a_{\rm h}$.
$a'_{\rm h,crit}$ depends on $c_{+}$ and $c_{-}$.
As $a'_{\rm h}$ approaches $a'_{\rm h,crit}$,
%$ \frac{dM_{\rm ADM}}{da'_{\rm H}} $
$ dM_{\rm ADM} / da'_{\rm h} $ tends to diverge. 
Since we obtain solutions numerically, it is nontrivial whether $ M_{\rm ADM} $ is bounded or not
from below. In particular, it is important to 
reveal the positivity of $ M_{\rm ADM} $.
However, since the energy conditions are not guaranteed~\cite{Eling},
we cannot prove it at present.

For $ M_{\rm ADM} $, the difference caused by the change of $c_{-}$ is not clear. 
We can define total energy $ M_{\rm TOT} $ by $ G_{\rm N} M_{\rm TOT} = G M_{\rm ADM} $
since the gravitational constant we feel is different from that in GR
as seen in Eq.~(\ref{Newton}). 
We also exhibit the relation $ a'_{\rm h} $-$M_{\rm TOT}$ in Fig.~\ref{a'-M} (b). 
This figure shows the differences caused by the change of $c_{-}$. 
$ M_{\rm TOT} $ decreases as $c_{-}$ increases as similar to $c_{+}$. 

If we contemplate these diagrams from a different viewpoint, 
we notice that the horizon radius of black holes in EA theory is larger than 
that of a Schwarzschild black hole for fixed $GM_{\rm TOT}$ (or $GM_{\rm ADM}$). 
Therefore, one might think that black holes in EA theory have larger entropy. 
However, since we have the Lorentz violating field, it is 
nontrivial to establish black hole thermodynamics~\cite{thermo,generalized}. 
Thus, the comparison of the black hole entropy, which is crucial to discuss the 
stability of black holes, belongs among our future tasks.

\begin{figure}[htbp]
\psfig{file=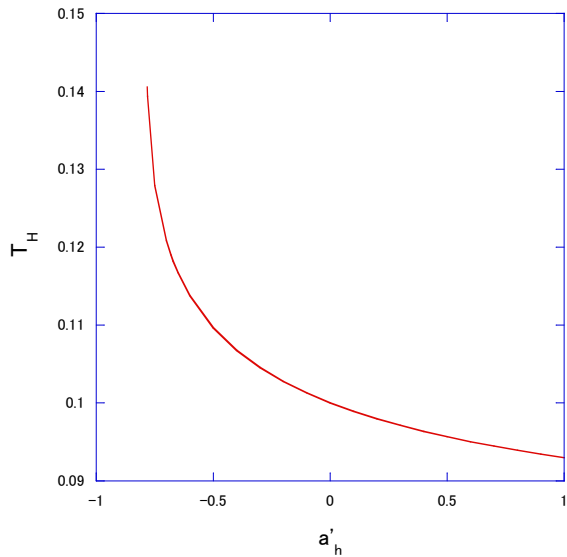,width=3in}
\caption{Hawking temperature $T_{\rm H}$ (normalized by $r_{\rm h}$) 
for $c_{+}=0.6$ and $c_{-}=0.1$ suggesting that $T_{\rm H}$ diverges for the solution at 
$a'_{\rm h,crit}$. 
 \label{temperature} }
\end{figure}

It is also important to reveal what happens at the critical point, $a'_{\rm h}=a'_{\rm h,crit}$. 
The key point is the factor $(-1+Na^{2})$ in the denominator in (\ref{eqs-a}). 
For $a'_{\rm h}=a'_{\rm h,crit}$, $(-1+Na^{2})$ becomes zero at finite $r$. 
Thus, solutions disappear. 
We also show the $a'_{\rm h}$ dependence of Hawking temperature 
$T_{\rm H}$ for $c_{+}=0.6$ and $c_{-}=0.1$ in Fig.~\ref{temperature}.
From this diagram, it is supposed that $T_{\rm H}$ diverges for the solution at 
$a'_{\rm h}=a'_{\rm h,crit}$. Thus, 
It is intriguing to consider an evaporation process of such black holes. 

%%%%%%%%%%%%%%%%%%%
%%%%%%%%%%%%%%%%%%%

%%%%%%%%%%%%%%%%%%%%%%%%%%%%%%%%
\subsection{ISCO of EA black hole}
%%%%%%%%%%%%%%%%%%%%%%%%%%%%%%%%
We shall turn to more realistic problems. 
We consider the possibility of distinguishing black holes in EA theory from 
Schwarzschild black hole by observation. In Ref.~\cite{NS}, the innermost 
stable circular orbit (ISCO) for neutron stars in EA theory was analyzed. 
The result is that the deviation from the Schwarzschild black hole is at most several percent. 
But this is not necessarily the case in the present situation, as shown below.
The differences occur since we have the freedom parameterized by $a'_{\rm h}$ 
and the Aether is not static. 
These facts will be important if we consider observations 
such as Constellation-X~\cite{X-ray}, which tracks the motion of individual elements 
near black holes. 

From an equation for timelike geodesics for a unit mass particle, 
we have effective potential $V$ as 
%%%%%%%%%%%%%%%%%%%%
\begin{eqnarray}
	V(r) = \frac{N}{B^2}\left(\frac{L^{2}}{r^{2}}+1\right), 
\end{eqnarray}
%%%%%%%%%%%%%%%%%%%%
where $L$ is the angular momentum normalized by the horizon radius. 

\begin{figure}[htbp]
\psfig{file=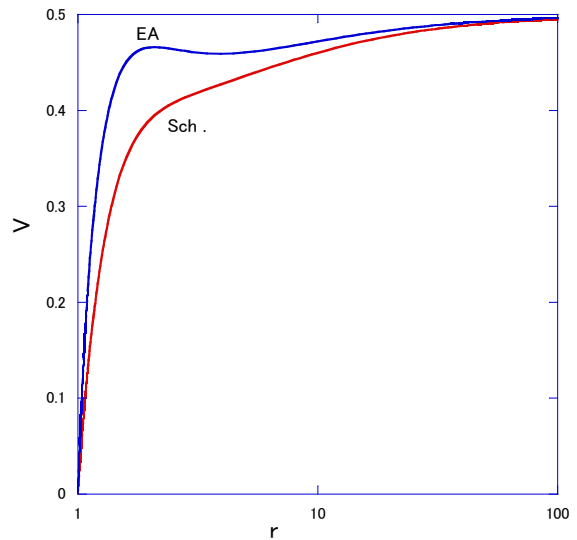,width=3in}
\caption{The potential $V$ for EA theory ($c_{+}=0.6$, $c_{-}=0.1$ and 
$a'_{\rm h}=0.78\simeq a'_{\rm h,crit}$) and for a Schwarzschild black hole where 
the angular momentum of the test particle $L$ (normalized by $r_{\rm h}$) 
is fixed by $L=1.5$. There is a potential minimum in EA theory while there is none 
for a Schwarzschild black hole. \label{comISCO} }
\end{figure}

We show the typical configurations of $V$ in Fig.~\ref{comISCO} for EA theory 
(with $c_{+}=0.6$, $c_{-}=0.1$ and $a'_{\rm h}=0.78\simeq a'_{\rm h,crit}$) and for 
GR (Schwarzschild black hole), 
where the angular momentum of the test particle $L$ is fixed as $L=1.5$. 
We find that a potential minimum exists even for $L=1.5$ in EA theory. 

\begin{figure}[htbp]
\psfig{file=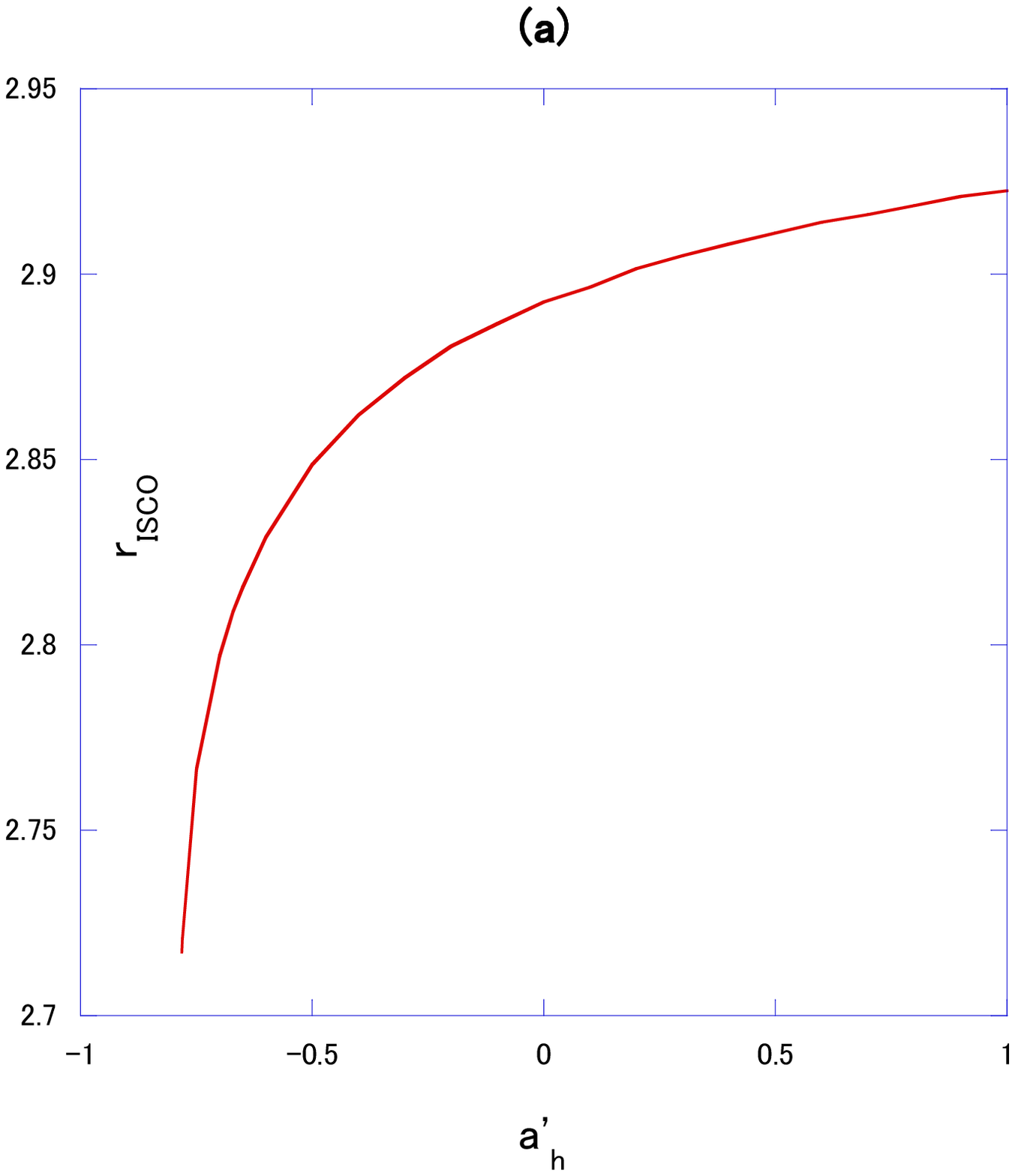,width=3in}
\psfig{file=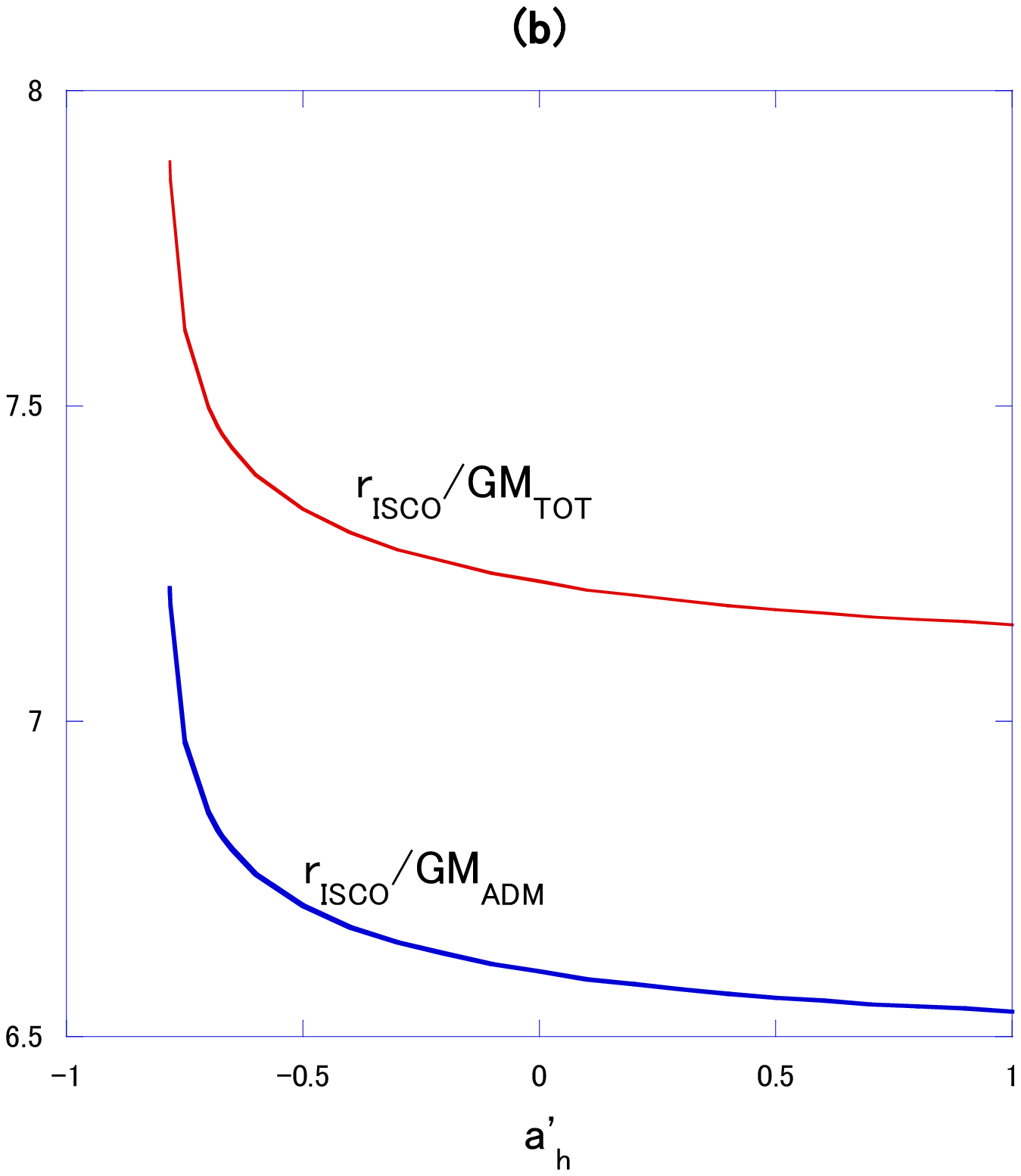,width=3in}
\caption{ The $a'_{\rm h}$ dependence of the innermost stable circular orbit (ISCO) for 
EA theory with $c_{+}=0.6$ and $c_{-}=0.1$. (a) ISCO normalized by $r_{\rm h}$. 
(b) ISCO normalized by $GM_{\rm TOT}$ and $GM_{\rm ADM}$. 
\label{ISCO} }
\end{figure}

We show the dependence of $r_{\rm ISCO}$ (normalized by $r_{\rm h}$) on $a'_{\rm h}$ in 
Fig.~\ref{ISCO} (a). 
Notice that $r_{\rm ISCO}=3$ for the Schwarzschild black hole.
Therefore, the difference is nearly $10 \%$ for $a'_{\rm h}\simeq a'_{\rm h,crit}$. 
It is also impressive to write the ISCO normalized by $GM_{\rm TOT}$ (or $GM_{\rm ADM}$), 
which is shown in Fig.~\ref{ISCO} (b). In this case, we can find the difference 
from the Schwarzschild black hole ($r_{\rm ISCO}/GM_{\rm ADM}=6$) is more than $20 \%$. 

Finally, let us comment on the parameter region of $(c_{+},c_{-})$ 
in which black hole solutions exist. 
We obtained solutions even for $c_{+}$, $c_{-}>1$, which seems to conflict 
with the previous results~\cite{BHs}. However, since we do not assume 
regularity at the spin-$0$ horizon against the case in \cite{BHs},
it is not inconsistent. The qualitative properties are same as in other 
parameter regions, although quantitative differences from Schwarzschild 
black hole become larger for large $(c_{+},c_{-})$ as we expect from 
Fig.~\ref{a'-M}. These features are same as in \cite{BHs} where the consistency 
with the weak gravity tests are not necessarily imposed. 

%%%%%%%%%%%%%%%%%%%%
%%%%%%%%%%%%%%%%%%%%

%%%%%%%%%%%%%%%%%%%%
%%%%%%%%%%%%%%%%%%%%
%%%%%%%%%%%%%%%%%%%%%%%%%%%%%%%%%%%
\section{Conclusion and discussion}
%%%%%%%%%%%%%%%%%%%%%%%%%%%%%%%%%%%

We have reanalyzed black hole solutions in EA theory while assuming 
that all the PPN parameters are the same as those for GR, resulting in two theoretical 
parameters $c_{+}$ and $c_{-}$. This is a main difference from the previous study~\cite{BHs}. 
As another difference, 
we do not assume regularity at the spin-$0$ horizon since this is inside the event horizon. 
Interestingly, we find $a'_{\rm h,crit}$ below which there is no regular black hole solution. 
Near $a'_{\rm h,crit}$, the deviation of black hole mass $M_{\rm TOT}$ 
(or $M_{\rm ADM}$) and ISCO $r_{\rm ISCO}$ from those for the Schwarzschild black hole 
become large. 

These results are instructive for other cases. 
If we consider the case with rotation, freedom of the vector field is added to 
(\ref{vector-setup}). Then, it also contributes the kinetic term of the vector field, 
enhancing the differences from the vacuum solution. 
This would also be true in other vector-tensor theories. 
For this reason, it is important to consider rotational black holes in vector-tensor 
theories, if we are to constrain them. 

Although we have revealed many properties of EA black holes, some important problems 
remain to be investigated. 
One is the positivity of the energy, which is necessary for 
the stability of the system. 
Related to this, to establish the black hole thermodynamics is also important. 
As a consistency check, we should also perform the linear perturbation for the 
black holes~\cite{foot1}. 

The other is whether or not regular spin-$0$ horizon happens as a result of 
gravitational collapse. In \cite{Garfinkle}, it is shown that regular spin-$0$ horizon happens 
if we consider a gravitational collapse of a massless scalar field. 
Thus, it is important to investigate this feature in a general case. 
It is also interesting to investigate the critical behavior of such a system~\cite{Choptuik}.
Of course, these are not problems particular only to EA theory but also issue confronting in 
more generic vector-tensor theories. Thus, it is desirable to investigate them 
in a unified way.

%%%%%%%%%%%%%%%%%%%%%%%%%%%%%%%%%%%%%%%%%%%
\acknowledgements
%%%%%%%%%%%%%%%%%%%%%%%%%%%%%%%%%%%%%%%%%%%
We would like to thank Kei-ichi Maeda for continuous encouragement. 
The numerical calculations were carried out on the Altix3700 BX2 at YITP, Kyoto University. 
This work is supported in part by a fund from the 21st Century COE Program (Holistic Research and 
Education Center for Physics of Self-Organizing Systems) at Waseda University.

\appendix
%%%%%%%%%%%%%%%%%%%%%%%%%%%%%%%%%%%%%%%%%%%
\section{basic equations for Einstein-Aether system}
%%%%%%%%%%%%%%%%%%%%%%%%%%%%%%%%%%%%%%%%%%%
The equation for $N$ is 
%%%%%%%%%%%%%%%%%%%%%
\begin{eqnarray}
&&\sum_{i=0}^{6}H_{i}a^{i}=0\ ,\label{quadratic}
\end{eqnarray}
%%%%%%%%%%%%%%%%%%%%%
where
%%%%%%%%%%%%%%%%%%%%%
\begin{eqnarray}
\hspace{-0.3cm}&&H_{6}=c_{+}[(3c_{-} - c_{+})N^{2} + 
2(3c_{-} + c_{+})rNN' + 
\nonumber  \\
\hspace{-0.3cm}&&(3c_{-} - c_{+})r^{2}N'^{2}\ , 
\nonumber  \\
\hspace{-0.3cm}&&H_{5}=2c_{+}rN[(3c_{-} + c_{+})N + (3c_{-} - c_{+})rN']a' \ ,
\nonumber  \\
\hspace{-0.3cm}&&H_{4}=-12(c_{-} + c_{+})B(r)^{2} + 
2[-3c_{-}(-2 + c_{+}) +   
\nonumber  \\
\hspace{-0.3cm}&&c_{+}(6+ c_{+})]N -2[3c_{-}(-2 + c_{+}) + (-6 + c_{+})c_{+}]rN' + 
\nonumber  \\
\hspace{-0.3cm}&&(3c_{-} - c_{+})c_{+}r^{2}N^{2}a'^{2}\ ,
\nonumber  \\
\hspace{-0.3cm}&&H_{3}=-2c_{+}(3c_{-} + c_{+})r^{2}N'a'\ ,
\nonumber  \\
\hspace{-0.3cm}&&H_{2}=-c_{+}a^{2}[-3c_{-} + c_{+} + 2(3c_{-} + c_{+})r^{2}Na'^{2}]\ ,
\nonumber  \\
\hspace{-0.3cm}&&H_{1}=-2c_{+}(3c_{-} + c_{+})ra'  \ , 
\nonumber  \\
\hspace{-0.3cm}&&H_{0}=(3c_{-} - c_{+})c_{+}r^{2}a'^{2}\ .
\nonumber
\end{eqnarray}
%%%%%%%%%%%%%%%%%%%%%
(\ref{quadratic}) is the quadratic equation for $N'$ (Notice $H_{6}$). 
If we solve (\ref{quadratic}) about $N'$, we obtain the equation which satisfies 
asymptotically flatness as 
%%%%%%%%%%%%%%%%%%%%%
\begin{eqnarray}
&&N'=\frac{ \sum_{i=0}^{3}h_{i}a^{i}+2\sqrt{3}\sqrt{X} }
{(3c_{-}-c_{+})c_{+}ra^{3}}\ ,  \label{eqs-N}  
\end{eqnarray}
%%%%%%%%%%%%%%%%%%%%%%
where 
%%%%%%%%%%%%%%%%%%%%%%%
\begin{eqnarray}
\hspace{-0.3cm}&& h_{3}=-c_{+}(3c_{-}+c_{+})N \ ,\ \  h_{2}=c_{+}(c_{+}-3c_{-})rNa'\ , 
\nonumber  \\
\hspace{-0.3cm}&& h_{1}=3c_{-}(c_{+}-2)+(c_{+}-6)c_{+}\ ,\ \ h_{0}=c_{+}(3c_{-}+c_{+})ra'\ ,
\nonumber  \\
\hspace{-0.3cm}&& X=[-3c_{-}^{2}(c_{+}-1)-(c_{+}-3)c_{+}^{2} +c_{-}c_{+}(6 -4c_{+} +
\nonumber \\
\hspace{-0.3cm}&&  c_{+}^{2})]a^{2}+c_{+}[(3c_{-}^{2}+2c_{-}c_{+}-c_{+}^{2})B^{2}+2c_{+}(c_{-}+c_{+}- 
\nonumber  \\
\hspace{-0.3cm}&& c_{-}c_{+})N]a^{4} +c_{-}c_{+}^{3}N^{2}a^{6}+c_{+}[3c_{-}^{2}(c_{+}-1)+
\nonumber  \\
\hspace{-0.3cm}&&c_{-}(c_{+}-4)c_{+}-c_{+}^{2}]raa' -c_{+}[3c_{-}^{2}(c_{+}-1)+
\nonumber  \\
\hspace{-0.3cm}&&c_{-}(c_{+}-2)c_{+}+c_{+}^{2}]rNa^{3}a'
+ c_{-}c_{+}^{3}r^{2}a'^{2} \ .\nonumber  
\end{eqnarray} 
%%%%%%%%%%%%%%%%%%%%%%%
Notice the denominator in (\ref{eqs-N}). For $c_{+}=3c_{-}$, 
we should use (\ref{quadratic}). 

The equation for $B$ is 
%%%%%%%%%%%%%%%%%%%%%%%
\begin{eqnarray}
\hspace{-1cm}&&B'=\frac{ B\sum_{i=0}^{8}g_{i}a^{i} }
{ Y }\ ,\label{eqs-B}
\end{eqnarray}
%%%%%%%%%%%%%%%%%%%%%%%%%%%%%%%%%
where 
%%%%%%%%%%%%%%%%%%%%%%%%%%%%%%%%%
\begin{eqnarray}
\hspace{-0.3cm}&&g_{8}=-c_{+}[3c_{-}^{2}(c_{+}-1)
+c_{-}(c_{+}-2)c_{+}+c_{+}^{2}]N \times 
\nonumber  \\
\hspace{-0.3cm}&&[(3c_{-}-c_{+})N^{2}+2(3c_{-}+c_{+})rNN'+(3c_{-}-c_{+})r^{2}N'^{2}] \ ,
\nonumber  \\
\hspace{-0.3cm}&&g_{7}=2c_{+}[3c_{-}^{2}(c_{+}-1)+c_{-}(c_{+}-2)c_{+}+c_{+}^{2}]rN^{2}\times
\nonumber  \\
\hspace{-0.3cm}&&[(3c_{-}+c_{+})N +(3c_{-}-c_{+})rN']a'\ ,
\nonumber  \\
\hspace{-0.3cm}&&g_{6}=-\{12[3c_{-}^{3}(c_{+}-1)+c_{-}(c_{+}-1)c_{+}^{2}+c_{+}^{3}+c_{-}^{2}c_{+}\times
\nonumber  \\
\hspace{-0.3cm}&&(4c_{+}-5)]B^{2}N+
[-c_{+}^{3}(12+c_{+})+c_{-}c_{+}^{2}(12-19c_{+}+ 
\nonumber  \\
\hspace{-0.3cm}&&3c_{+}^{2}) +9c_{-}^{3}(4-9c_{+}+5c_{+}^{2})+3c_{-}^{2}c_{+}(20-33c_{+}+8c_{+}^{2})]N^{2}
\nonumber  \\
\hspace{-0.3cm}&&+4[-3c_{+}^{3}+c_{-}c_{+}^{2}(3-9c_{+}-2c_{+}^{2})+9c_{-}^{3}(1-3c_{+}+2c_{+}^{2})+
\nonumber  \\
\hspace{-0.3cm}&&3c_{-}^{2}c_{+}(5-12c_{+}+4c_{+}^{2})]rNN'+c_{+}[27c_{-}^{3}(c_{+}-1)-
\nonumber  \\
\hspace{-0.3cm}&&21c_{-}^{2}c_{+}+c_{+}^{3}+c_{-}c_{+}^{2}(7+5c_{+})]r^{2}N'^{2}
+c_{+}[-9c_{-}^{3}(c_{+}-1)+
\nonumber  \\
\hspace{-0.3cm}&&3c_{-}^{2}c_{+}+c_{-}(-5+c_{+})c_{+}^{2}+c_{+}^{3}]r^{2}N^{3}a'^{2}\} \ ,
\nonumber  \\
\hspace{-0.3cm}&&g_{5}=-2c_{+}rN\{ [27c_{-}^{3}(c_{+}-1)-c_{+}^{3}-c_{-}c_{+}^{2}(13+ 5c_{+})+
\nonumber  \\
\hspace{-0.3cm}&&3c_{-}^{2}c_{+}(-13+6c_{+})]N+2[18c_{-}^{3}(c_{+}-1)+3c_{-}^{2}(-5+
\nonumber  \\ 
\hspace{-0.3cm}&&c_{+})c_{+}+c_{+}^{3}+c_{-}c_{+}^{2}(4+3c_{+})]rN' \} a' \ ,
\nonumber  \\ 
\hspace{-0.3cm}&&g_{4}=12[9c_{-}^{3}(c_{+}-1)-c_{+}^{3}+c_{-}c_{+}^{2}(-11+3c_{+})+
c_{-}^{2}c_{+}\times
\nonumber  \\
\hspace{-0.3cm}&&(-19+12c_{+})]B^{2}+[c_{+}^{3}(12+c_{+})+9c_{-}^{3}(12-19c_{+}+
\nonumber  \\
\hspace{-0.3cm}&&7c_{+}^{2})+c_{-}c_{+}^{2}(132-65c_{+}+9c_{+}^{2})+3c_{-}^{2}c_{+}(76-79c_{+}+
\nonumber  \\
\hspace{-0.3cm}&&16c_{+}^{2})]N+2[-(c_{+}-6)c_{+}^{3}+c_{-}c_{+}^{2}(66 -31c_{+}-5c_{+}^{2})+
\nonumber  \\
\hspace{-0.3cm}&&27c_{-}^{3}(2-3c_{+}+c_{+}^{2})+3c_{-}^{2}c_{+}(38-37c_{+}+6c_{+}^{2})]rN'-
\nonumber  \\
\hspace{-0.3cm}&&c_{+}[45c_{-}^{3}(c_{+}-1)+3c_{+}^{3}+3c_{-}^{2}c_{+}(-13+4c_{+})+
\nonumber  \\
\hspace{-0.3cm}&&c_{-}c_{+}^{2}(9+7c_{+})]r^{2}N^{2}a'^{2}\ ,
\nonumber  \\
\hspace{-0.3cm}&&g_{3}=-2c_{+}r\{ [9c_{-}^{3}(c_{+}-1)+c_{+}^{3} +c_{-}c_{+}^{2}(1+c_{+})+
\nonumber  \\
\hspace{-0.3cm}&&3c_{-}^{2}c_{+}(-3+2c_{+})]N+
[-27c_{-}^{3}(c_{+}-1)+c_{+}^{3}+
\nonumber  \\
\hspace{-0.3cm}&&3c_{-}^{2}c_{+}(5+2c_{+})+c_{-}c_{+}^{2}(-11+5c_{+})]rN'     \} a'\ ,
\nonumber  \\
\hspace{-0.3cm}&&g_{2}=-c_{+}\{ 27c_{-}^{3}(c_{+}-1)+c_{+}^{3}+c_{-}c_{+}^{2}(-17+5c_{+})+
\nonumber  \\
\hspace{-0.3cm}&&3c_{-}^{2}c_{+}(-15+8c_{+})+[-63c_{-}^{3}(c_{+}-1)+3c_{+}^{3}+
\nonumber  \\
\hspace{-0.3cm}&&3c_{-}^{2}c_{+}(11+4c_{+})+c_{-}c_{+}^{2}(-27+11c_{+})]r^{2}Na'^{2}\}\ ,
\nonumber  \\
\hspace{-0.3cm}&&g_{1}=-2c_{+}[-27c_{-}^{3}(c_{+}-1)+3c_{-}^{2}(13-6c_{+})c_{+}+c_{+}^{3}+
\nonumber  \\
\hspace{-0.3cm}&&c_{-}c_{+}^{2}(13+ 5c_{+})]ra'  \ ,
\nonumber  \\
\hspace{-0.3cm}&&g_{0}=-c_{+}[27c_{-}^{3}(c_{+}-1)-21c_{-}^{2}c_{+}+
\nonumber  \\
\hspace{-0.3cm}&&c_{+}^{3}+c_{-}c_{+}^{2}(7+5c_{+})]r^{2}a'^{2}\ ,
\nonumber  \\
\hspace{-0.3cm}&&Y=12[c_{-}^{2}(c_{+}-1)+c_{-}(c_{+}-2)c_{+}-
c_{+}^{2}]ra^{2}\times  
\nonumber  \\
\hspace{-0.3cm}&&\{ 2[-3c_{-}(c_{+}-1)+c_{+}]Na^{2}+
\nonumber  \\
\hspace{-0.3cm}&&[3c_{-}(c_{+}-1)+c_{+}](N^{2}a^{4}+1)\} \ .\nonumber  
\end{eqnarray}
%%%%%%%%%%%%%%%%%%%%%%%%%%%%%%%%%

The equation for $a$ is 
%%%%%%%%%%%%%%%%%%%%%%%%%%%%%%%%%
\begin{eqnarray}
&&a''=\frac{ \sum_{i=0}^{13}f_{i}a^{i} }{ c_{+}r(-1+Na^{2})Y }  \ , 
\label{eqs-a}
\end{eqnarray} 
%%%%%%%%%%%%%%%%%%%%%%%%%%%%%%%%%
where 
%%%%%%%%%%%%%%%%%%%%%%%%%%%%%%%%%
\begin{eqnarray}
\hspace{-0.3cm}&&f_{13}=2c_{-}c_{+}^{4}N^{3}[(3c_{-}-c_{+})N^{2} +2(3c_{-}+c_{+})rNN'+
\nonumber  \\
\hspace{-0.3cm}&&(3c_{-}-c_{+})r^{2}N'^{2}]\ , 
\nonumber  \\
\hspace{-0.3cm}&&f_{12}=4c_{-}c_{+}^{4}rN^{4}[(3c_{-}+c_{+})N+(3c_{-}-c_{+})rN']a'\ ,
\nonumber \\
\hspace{-0.3cm}&&f_{11}=c_{+}^{2}N^{2}\{ -24c_{-}c_{+}(c_{-}+c_{+})B^{2}N+[9c_{-}^{3}(c_{+}-
\nonumber \\
\hspace{-0.3cm}&&1)-7c_{+}^{3}+c_{-}c_{+}^{2}(17+7c_{+}) -3c_{-}^{2}c_{+}(-5+8c_{+})]N^{2}+
\nonumber  \\
\hspace{-0.3cm}&&2[9c_{-}^{3}(c_{+}-1)+c_{-}(13 -5c_{+})c_{+}^{2}+7c_{+}^{3}-3c_{-}^{2}c_{+}(1+
\nonumber  \\
\hspace{-0.3cm}&&4c_{+})]rNN'+[9c_{-}^{3}(c_{+}-1)+3c_{-}^{2}(5-12c_{+})c_{+}-7c_{+}^{3}+
\nonumber  \\
\hspace{-0.3cm}&&c_{-}c_{+}^{2}(17+3c_{+})]r^{2}N'^{2}+
2c_{-}(3c_{-}-c_{+})c_{+}^{2}r^{2}N^{3}a'^{2}\} \ ,
\nonumber  \\
\hspace{-0.3cm}&&f_{10}=c_{+}^{2}rN^{2}\{ [9c_{-}^{3}(c_{+}-1)+15c_{+}^{3}-
3c_{-}^{2}c_{+}(9+4c_{+})-
\nonumber  \\
\hspace{-0.3cm}&&c_{-}c_{+}^{2}(3+5c_{+})]N^{2}-16c_{+}[-2c_{-}c_{+}+c_{+}^{2}+c_{-}^{2}(-3+
\nonumber  \\
\hspace{-0.3cm}&&6c_{+})]rNN'+[-9c_{-}^{3}(c_{+}-1)+3c_{-}^{2}c_{+}+
\nonumber  \\
\hspace{-0.3cm}&&c_{-}(-5+c_{+})c_{+}^{2}+c_{+}^{3}]r^{2}N'^{2}\} a'\ ,
\nonumber  \\
\hspace{-0.3cm}&&f_{9}=c_{+}N\bigl( -12[3c_{-}^{3}(c_{+}-1)+c_{-}(3-7c_{+})c_{+}^{2}+3c_{+}^{3}-
\nonumber  \\
\hspace{-0.3cm}&&c_{-}^{2}c_{+}(3+4c_{+})]B^{2}N+
2[(18-13c_{+})c_{+}^{3}+3c_{-}^{2}c_{+}(-6+
\nonumber  \\
\hspace{-0.3cm}&&7c_{+})+c_{-}c_{+}^{2}(18-37c_{+}+3c_{+}^{2})-9c_{-}^{3}(2-5c_{+}+
\nonumber  \\
\hspace{-0.3cm}&&3c_{+}^{2})]rNN'+6c_{+}[-3c_{-}^{3}(c_{+}-1)-7c_{-}c_{+}^{2}+c_{+}^{3}+
\nonumber  \\
\hspace{-0.3cm}&&c_{-}^{2}c_{+}(-5+9c_{+})]r^{2}N'^{2}-3c_{+}[3c_{-}^{3}(c_{+}-1)+
\nonumber  \\
\hspace{-0.3cm}&&c_{-}(-5+c_{+})c_{+}^{2}+3c_{+}^{3}+c_{-}^{2}c_{+}(-11+20c_{+})]r^{2}N^{3}a'^{2}- 
\nonumber  \\
\hspace{-0.3cm}&&2N^{2}\{ 2[9c_{-}^{3}(c_{+}-1)^{2} -c_{+}^{3}(3+5c_{+})+c_{-}c_{+}^{2}(3+10c_{+}+
\nonumber  \\
\hspace{-0.3cm}&&2c_{+}^{2})-3c_{-}^{2}c_{+}(-5+c_{+}+3c_{+}^{2})]-c_{+}[-9c_{-}^{3}(c_{+}-1)+
\nonumber  \\
\hspace{-0.3cm}&&3c_{-}^{2}c_{+}+c_{-}(-5+c_{+})c_{+}^{2}+c_{+}^{3}]r^{3}N'a'^{2}\}\bigr)\ ,
\nonumber  \\
\hspace{-0.3cm}&&f_{8}=c_{+}rNa'\{ 12[3c_{-}^{3}(c_{+}-1) +c_{-}(c_{+}-1)c_{+}^{2} +c_{+}^{3}+
\nonumber  \\
\hspace{-0.3cm}&&c_{-}^{2}c_{+}(4c_{+}-5)]B^{2}N-2[c_{+}^{3}(-6+7c_{+})+c_{-}c_{+}^{2}(6-
\nonumber  \\
\hspace{-0.3cm}&&23c_{+}+c_{+}^{2})+9c_{-}^{3}(2-5c_{+}+3c_{+}^{2})+3c_{-}^{2}c_{+}(10-25c_{+}+
\nonumber  \\
\hspace{-0.3cm}&&12c_{+}^{2})]N^{2}-12[3c_{-}^{3}(c_{+}-1)^{2}-c_{+}^{3} +c_{-}c_{+}^{2}(1+10c_{+})+
\nonumber  \\
\hspace{-0.3cm}&&c_{-}^{2}c_{+}(5+4c_{+}-13c_{+}^{2})]rNN'+4c_{-}c_{+}[9c_{-}^{2}(c_{+}-1)-
\nonumber  \\
\hspace{-0.3cm}&&6c_{-}c_{+}+c_{+}^{2}(3+c_{+})]r^{2}N'^{2}+c_{+}[-9c_{-}^{3}(c_{+}-1)+
\nonumber  \\
\hspace{-0.3cm}&&3c_{-}^{2}c_{+}+c_{-}(-5+c_{+})c_{+}^{2}+c_{+}^{3}]r^{2}N^{3}a'^{2}\} \ ,
\nonumber  \\
\hspace{-0.3cm}&&f_{7}=c_{+}\bigl( 24[3c_{-}^{3}(c_{+}-1)+c_{-}(3-4c_{+})c_{+}^{2}+3c_{+}^{3}-
\nonumber  \\
\hspace{-0.3cm}&&c_{-}^{2}c_{+}(3+c_{+})]B^{2}N+2c_{+}[27c_{-}^{3}(c_{+}-1)+c_{+}^{2}(-48+
\nonumber  \\
\hspace{-0.3cm}&&5c_{+})+c_{-}c_{+}(-96+47c_{+}+c_{+}^{2})+3c_{-}^{2}(-16+5c_{+}+
\nonumber  \\
\hspace{-0.3cm}&&4c_{+}^{2})]rNN'+[c_{+}^{3}(12+c_{+})-c_{-}c_{+}^{2}(12-13c_{+}+c_{+}^{2})+
\nonumber  \\
\hspace{-0.3cm}&&9c_{-}^{3}(-4+3c_{+}+c_{+}^{2})-3c_{-}^{2}c_{+}(20-13c_{+}+8c_{+}^{2})]r^{2}N'^{2}-
\nonumber  \\
\hspace{-0.3cm}&&4c_{+}[-9c_{-}^{3}(c_{+}-1)+2c_{+}^{3}+c_{-}c_{+}^{2}(29+c_{+})-
\nonumber  \\
\hspace{-0.3cm}&&6c_{-}^{2}c_{+}(-6+7c_{+})]r^{2}N^{3}a'^{2}+2N^{2}\{ -9c_{+}^{3}(2+c_{+})+
\nonumber  \\
\hspace{-0.3cm}&&c_{-}c_{+}^{2}(-18+21c_{+}+c_{+}^{2})+9c_{-}^{3}(2-5c_{+}+3c_{+}^{2})- 
\nonumber  \\
\hspace{-0.3cm}&&3c_{-}^{2}c_{+}(-6+5c_{+}+4c_{+}^{2})+c_{+}[45c_{-}^{3}(c_{+}-1)+c_{+}^{3}+
\nonumber  \\
\hspace{-0.3cm}&&3c_{-}^{2}c_{+}(-11+2c_{+})+c_{-}c_{+}^{2}(13+5c_{+})]r^{3}N'a'^{2}\} \bigr) \ ,
\nonumber  \\
\hspace{-0.3cm}&&f_{6}=c_{+}ra'\{ -48c_{-}[3c_{-}^{2}(c_{+}-1)+(c_{+}-3)c_{+}^{2} +
\nonumber  \\
\hspace{-0.3cm}&&2c_{-}c_{+}(-3+2c_{+})]B^{2}N-4c_{+}[-27c_{-}^{3}(c_{+}-1)+
\nonumber  \\
\hspace{-0.3cm}&&6c_{-}^{2}(10-7c_{+})c_{+}+4c_{+}^{3}+c_{-}c_{+}^{2}(37+c_{+})]N^{2}+
\nonumber  \\
\hspace{-0.3cm}&&8[c_{+}^{3}(3+2c_{+})+9c_{-}^{3}(-1+c_{+}^{2}) +c_{-}c_{+}^{2}(-3+8c_{+}
\nonumber  \\
\hspace{-0.3cm}&&+2c_{+}^{2})-3c_{-}^{2}c_{+}(5-2c_{+}+3c_{+}^{2})]rNN'-c_{+}[27c_{-}^{3}(c_{+}-
\nonumber  \\
\hspace{-0.3cm}&&1)-21c_{-}^{2}c_{+}+c_{+}^{3}+c_{-}c_{+}^{2}(7+5c_{+})]r^{2}N'^{2}+
\nonumber  \\
\hspace{-0.3cm}&&2c_{+}[27c_{-}^{3}(c_{+}-1)+c_{+}^{3}+3c_{-}^{2}c_{+}(-7+2c_{+})+
\nonumber  \\
\hspace{-0.3cm}&&c_{-}c_{+}^{2}(7+3c_{+})]r^{2}N^{3}a'^{2}\} \ ,
\nonumber  \\
\hspace{-0.3cm}&&f_{5}=-2\bigl( 18[(c_{+}-4)c_{+}^{3}+c_{-}^{2}c_{+}(-12+7c_{+})-
c_{-}c_{+}^{2}\times
\nonumber  \\
\hspace{-0.3cm}&&(12-5c_{+}+c_{+}^{2})+c_{-}^{3}(-4+3c_{+}+c_{+}^{2})]B^{2}+
\nonumber  \\
\hspace{-0.3cm}&&[-c_{+}^{3}(-72+30c_{+}+c_{+}^{2})+9c_{-}^{3}(8-10c_{+}+c_{+}^{2}+c_{+}^{3})+
\nonumber  \\
\hspace{-0.3cm}&&c_{-}c_{+}^{2}(216-150c_{+}+23c_{+}^{2}+c_{+}^{3})+3c_{-}^{2}c_{+}(72-70c_{+}+
\nonumber  \\
\hspace{-0.3cm}&&11c_{+}^{2}+2c_{+}^{3})]rN'+c_{+}[(6-5c_{+})c_{+}^{3}+c_{-}c_{+}^{2}(66- 
\nonumber  \\
\hspace{-0.3cm}&&119c_{+}-7c_{+}^{2})+27c_{-}^{3}(2-3c_{+}+c_{+}^{2})+3c_{-}^{2}c_{+}(38-
\nonumber  \\
\hspace{-0.3cm}&&65c_{+}+40c_{+}^{2})]r^{2}N^{2}a'^{2}+N\{ -2c_{+}^{3}(-36+9c_{+}+c_{+}^{2})+
\nonumber  \\
\hspace{-0.3cm}&&18c_{-}^{3}(4-5c_{+}+c_{+}^{3})-c_{-}c_{+}^{2}(-216+126c_{+}-16c_{+}^{2}+
\nonumber  \\
\hspace{-0.3cm}&&c_{+}^{3})-3c_{-}^{2}c_{+}(-72+66c_{+}-6c_{+}^{2}+c_{+}^{3})+c_{+}^{2}[63c_{-}^{3}(c_{+}-
\nonumber  \\
\hspace{-0.3cm}&&1)-45c_{-}^{2}c_{+}+c_{+}^{3}+c_{-}c_{+}^{2}(19+c_{+})]r^{3}N'a'^{2}\} \bigr) \ ,
\nonumber  \\
\hspace{-0.3cm}&&f_{4}=-2c_{+}ra'\{ -6[9c_{-}^{3}(c_{+}-1)-c_{+}^{3}+c_{-}c_{+}^{2}(3c_{+}-
\nonumber  \\
\hspace{-0.3cm}&&11)+c_{-}^{2}c_{+}(-19+12c_{+})]B^{2} +[(6-7c_{+})c_{+}^{3}+c_{-}c_{+}^{2}\times
\nonumber  \\
\hspace{-0.3cm}&&(66 -73c_{+}-7c_{+}^{2}) +9c_{-}^{3}(6-11c_{+}+5c_{+}^{2})+3c_{-}^{2}c_{+}\times
\nonumber  \\
\hspace{-0.3cm}&&(38- 55c_{+}+22c_{+}^{2})]N+6[-c_{+}^{3}+c_{-}^{2}c_{+}(-19+8c_{+})+
\nonumber  \\
\hspace{-0.3cm}&&c_{-}c_{+}^{2}(-11+2c_{+}+c_{+}^{2})+3c_{-}^{3}(-3+ 2c_{+}+c_{+}^{2})]rN'+
\nonumber  \\
\hspace{-0.3cm}&&2c_{-}c_{+}[27c_{-}^{2}(c_{+}-1)-18c_{-}c_{+}-(-9+c_{+})c_{+}^{2}]r^{2}N^{2}a'^{2}\}\ ,
\nonumber  \\
\hspace{-0.3cm}&&f_{3}=c_{+}\{ -36c_{-}^{3}-84c_{-}^{2}c_{+}+27c_{-}^{3}c_{+}-60c_{-}c_{+}^{2}+
\nonumber  \\
\hspace{-0.3cm}&&39c_{-}^{2}c_{+}^{2}+9c_{-}^{3}c_{+}^{2}-12c_{+}^{3}+13c_{-}c_{+}^{3}+c_{+}^{4}-c_{-}c_{+}^{4}+
\nonumber  \\
\hspace{-0.3cm}&&2[4c_{+}^{3}(3+c_{+})+18c_{-}^{3}(6-7c_{+}+c_{+}^{2})+c_{-}c_{+}^{2}(132-
\nonumber  \\
\hspace{-0.3cm}&&110c_{+}+3c_{+}^{2})+3c_{-}^{2}c_{+}(76-80c_{+}+31c_{+}^{2})]r^{2}Na'^{2}-
\nonumber  \\
\hspace{-0.3cm}&&2c_{+}[-27c_{-}^{3}(c_{+}-1)+c_{+}^{3}+3c_{-}^{2}c_{+}(5+2c_{+})+
\nonumber  \\
\hspace{-0.3cm}&&c_{-}c_{+}^{2}(-11+5c_{+})]r^{3}N'a'^{2}\} \ ,
\nonumber  \\
\hspace{-0.3cm}&&f_{2}=c_{+}ra'\{ c_{+}^{4}+c_{-}c_{+}^{2}(144-41c_{+}-7c_{+}^{2})+
\nonumber  \\
\hspace{-0.3cm}&&9c_{-}^{3}(16-19c_{+}+3c_{+}^{2})+3c_{-}^{2}c_{+}(96-71c_{+}+12c_{+}^{2})-
\nonumber  \\
\hspace{-0.3cm}&&2c_{+}[-45c_{-}^{3}(c_{+}-1)+c_{+}^{3}+3c_{-}^{2}c_{+}(9+2c_{+})+
\nonumber  \\
\hspace{-0.3cm}&&c_{-}c_{+}^{2}(-17+3c_{+})]r^{2}Na'^{2}\} \ ,
\nonumber  \\
\hspace{-0.3cm}&&f_{1}=-c_{+}[(-36+c_{+})c_{+}^{3}+9c_{-}^{3}(12 -13c_{+}+c_{+}^{2})+
\nonumber  \\
\hspace{-0.3cm}&&c_{-}c_{+}^{2}(36-35c_{+}+11c_{+}^{2})+
\nonumber  \\
\hspace{-0.3cm}&&3c_{-}^{2}c_{+}(60-51c_{+}+20c_{+}^{2})]r^{2}a'^{2}\ ,
\nonumber  \\
\hspace{-0.3cm}&&f_{0}=-c_{+}^{2}[27c_{-}^{3}(c_{+}-1)-21c_{-}^{2}c_{+}+c_{+}^{3} +
\nonumber  \\
\hspace{-0.3cm}&&c_{-}c_{+}^{2}(7+5c_{+})]r^{3}a'^{3} \ .\nonumber  
\end{eqnarray}
%%%%%%%%%%%%%%%%%%%%%%%%%%%%%%%%%

If we remove $N'$ from (\ref{eqs-B}) and (\ref{eqs-a}), we can write them as 
the form in (\ref{basiceq2}) and (\ref{basiceq3}). Since it is too tedious, 
we do not perform it. From (\ref{eqs-N}) to (\ref{eqs-a}), we obtain the 
asymptotic form for $r\to\infty$ as 
%%%%%%%%%%%%%%%%
\begin{eqnarray}
&&N(r)=B(\infty )^{2}+\frac{N_{1}}{r}+\cdots ,\label{asym-N}\\
&&B(r)=B(\infty )+\frac{B_{1}}{r^{2}}+\cdots ,\label{asym-B}\\
&&a(r)=\frac{1}{B(\infty )}+\frac{a_{1}}{r}+\cdots \label{asym-a},
\end{eqnarray}
%%%%%%%%%%%%%%%%
where $N_{1}$, $B_{1}$ and $a_{1}$ are constants. 

We should be careful about $(-1+Na^{2})$ in the denominator in (\ref{eqs-a}) since 
(\ref{flat}) and (\ref{flat3}) show that $(-1+Na^{2})$ asymptotically approaches zero. 
However, since $(-1+Na^{2})\propto 1/r$ for $r\to\infty$, this is canceled by $r$ 
in the denominator in (\ref{eqs-a}). 
%%%%%%%%%%%%%%%%%%%%%%%%%%%%%%%%%%%%%%%%%%%%%%%%%%%%%%%%%%%%%%%%%%%%%%%%

%%%%%%%%%%%%%%%%%%%%%%%%%%%%%%%%%%%%%%%%%%%%%%%%%%%%%%%%%%%%%%%%%%%%%%%%


\begin{thebibliography}{99}
\bibitem{foot1}
As for the solution with static Aether that describes outside the star, stability 
against spherically symmetric perturbation 
has been established in \cite{Seifert} under the condition (\ref{sonic}).
\end{thebibliography}
\end{document}